\documentclass[twocolumn,prl,floatfix,showpacs]{revtex4}

\usepackage{graphicx}
\usepackage{bm}
\begin{document}

\title{Atom interferometry with trapped Fermi gases}
\author{G. Roati, E. de Mirandes, F. Ferlaino, H. Ott, G. Modugno}
\author{M. Inguscio}
\affiliation{LENS and Dipartimento di Fisica, Universit\`a di Firenze, and INFM\\
 Via Nello Carrara 1, 50019 Sesto Fiorentino, Italy }

\begin{abstract}
We realize an interferometer with an atomic Fermi gas trapped in
an optical lattice under the influence of gravity. The
single-particle interference between the eigenstates of the
lattice results in macroscopic Bloch oscillations of the sample.
The absence of interactions between fermions allows a
time-resolved study of many periods of the oscillations, leading
to a sensitive determination of the acceleration of gravity. The
experiment proves the superiorness of non interacting fermions
with respect to bosons for precision interferometry, and offers a
way for the measurement of forces with microscopic spatial
resolution.
\end{abstract}
\date{\today}
\maketitle

In recent years ultracold atomic gases have been successfully
employed to perform high precision measurements with
interferometric methods \cite{chu,gradients,pritchard}. The
sensitivity of an interferometer usually increases with the
brightness of the source. The advent of  Bose-Einstein condensates
(BEC) was expected to produce in atom interferometry the same
dramatic progress faced by photon interferometry after the
invention of laser. A BEC is the brightest atom source with all
the particles in the same quantum state, hence leading to an
increase of the contrast of the interference signal
\cite{ketterle,anderson,phillips,pritchard,ketterleb}. However,
differently from photons, atoms interact and this can dramatically
affect an interferometric measurement, giving rise to a shift or
decay of the signal. To avoid this problem high-precision
interferometry in combination with BECs has been performed so far
only with samples in free fall where interactions are weaker, but
the observation times are limited \cite{pritchard}. On the other
hand, collisions are suppressed in an ultracold sample composed of
identical fermions. However, the same Pauli principle that forbids
collisions also limits the phase-space density of fermions to
unity. The question is now whether this constitutes an obstacle to
precise interferometry.

In the present work we observe the interference of fermionic atoms
trapped in the sites of a one-dimensional vertical optical
lattice. The gravitational energy difference between the
eigenstates of the system results in a periodic oscillation of the
macroscopic interference pattern, also known as Bloch oscillations
\cite{bloch}. The absence of interactions between atoms allows the
oscillations to proceed with high contrast for many periods. We
repeat the same experiment with a BEC: in this case the
interference shows a lower effective contrast and decays rapidly
because of interactions. We show how interferometry with trapped
fermions gives the possibility of a sensitive measurement of
forces with spatial resolution at the micrometer scale.

We employ fermionic $^{40}$K atoms, which are brought to quantum
degeneracy by sympathetic cooling with bosonic $^{87}$Rb atoms
\cite{roati}. Both species are held in a harmonic magnetic
potential and have a horizontal cigar shape. The bosons are
completely removed from the trap at the end of the evaporation,
leaving a pure Fermi gas of about 3$\times10 ^4$ atoms
spin-polarized in the $F$=9/2, $m_F$=9/2 state. The typical
temperature is $T$=0.3$~T_F$, where $T_F$=330~nK is the Fermi
temperature. We then switch on adiabatically a lattice created by
a retroreflected laser beam aligned along the vertical direction.
The wavelength of the lattice laser is far detuned to the red of
the optical atomic transitions ($\lambda$=873~nm) to avoid photon
scattering. The depth of the potential can be adjusted in the
range $U$=1-4~$E_R$, where
$E_R$=$h^2/2m\lambda^2$=$k_B\times$310~nK is the recoil energy. At
these low temperatures the atoms are loaded mostly in the first
Bloch band of the lattice. In the horizontal directions they are
confined by the gaussian profile of the lattice beams. Using
different intensities for the two beams we obtain a radial trap
depth of about 10~$E_R$, with a typical trapping frequency of
$2\pi\times30$~s$^{-1}$.

\begin{figure*}[t]
\includegraphics[width=17cm]{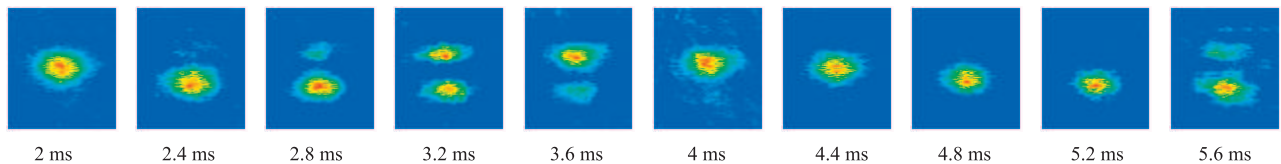}
\caption{Evolution of the interference pattern of the Fermi gas
for increasing holding times in the vertical lattice. The spatial
distribution of the cloud detected after 8 ms of free expansion
reflects the momentum distribution in the trap at the time of
release. } \label{fig2}
\end{figure*}
\begin{figure}
\includegraphics[width=8.5cm,clip=]{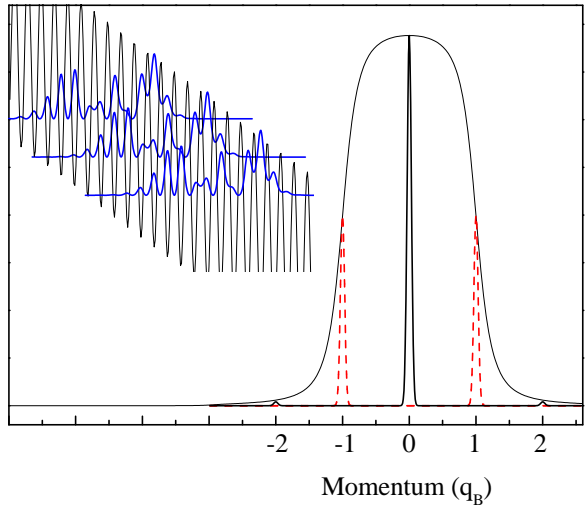} \caption{Momentum distribution
of a coherent superposition of Wannier-Stark states. Although the
single state fills completely the first Brillouin zone (thin
line), the interference of several states gives narrow momentum
peaks. Shown are the cases of a phase difference between
successive states of $\Delta\phi$=0 (continuous line) and
$\Delta\phi$=$\pi$ (dashed line). The inset shows the square of
Wannier-Stark wavefunctions calculated for $^{40}$K atoms in a
lattice with $U$=2$E_R$ subjected to gravity (for clarity, the
states shown are separated by four lattice sites).} \label{fig1}
\end{figure}
To observe the interference we suddenly switch off the magnetic
trap and let the atoms evolve in the lattice plus the
gravitational potential. The energy spectrum of such combined
potential is the well known Wannier-Stark ladder of states
\cite{wannier}, which are equally spaced by $\Delta
E$=$mg\lambda/2$. For the typical lattice depth $U$=2$E_R$ each of
the Wannier-Stark states extends over about ten lattice sites. The
vertical distribution of the atomic cloud has a gaussian envelope
with a fullwidth at 1/$e^2$ of 55~$\mu$m, and therefore extends
over about 125 sites \cite{width}. Radially, the atoms occupy the
harmonic oscillator states of the trap, with a fullwidth of
200~$\mu$m. Each of the atoms in the cloud is therefore prepared
in a coherent superposition of several Wannier-Stark states. Due
to their energy difference, neighboring states evolve in time with
a phase difference $\Delta\phi$= $\Delta E t/\hbar$, and their
interference pattern is periodic in time, with a period
$T_B$=$h/\Delta E$.

In particular, the interference of Wannier-Stark states results in
equally spaced peaks in momentum space that move with constant
velocity $\dot{q}$=$mg$. The peaks spacing is the inverse of the
spatial period of the lattice, and can be written as 2$q_B$, where
$q_B$=$h/\lambda$ is the Bragg momentum. Therefore only one or two
peaks appear at the same time in the first Brillouin zone of the
lattice $[-q_B,+q_B]$, as shown in Fig~\ref{fig1}.

To study the momentum distribution in the trap, we release the
cloud from the lattice, thus stopping the evolution of the
interference pattern at a given time. We then probe the cloud by
absorption imaging after a 8-ms ballistic expansion, which maps
the initial momentum distribution into a position distribution.
Actually, the lattice depth is lowered to zero in about 50~$\mu$s,
a time scale longer than the oscillation period of the atoms in
each lattice well. The adiabatic release allows to study the
evolution of the momentum in the first Brillouin zone.
Fig.~\ref{fig2} shows the time-evolution in $q$ space detected in
the experiment. We can clearly see the vertical motion of the peak
of the distribution, initially centered in $q$=0 at $t$=2~ms. It
gradually disappears as it reaches the lower edge of the Brillouin
zone at $t$=2.8~ms, while a second peak builds up at the upper
edge and then scans the whole Brillouin zone as the first one. The
periodicity of this interference pattern amounts to about 2.3~ms,
in agreement with the expected $T_B$=$2h/mg\lambda$.

A quantitative description of the observations becomes
particularly easy in a semiclassical approach. Here the atomic
cloud is described as a single wavepacket that moves uniformly in
$q$ space under the influence of gravity and is gradually
reflected each time it reaches the lower band edge. This
phenomenon are the well known Bloch oscillations
\cite{bloch,review}, which have been studied for a variety of
systems including cold atoms in accelerated horizontal lattices
\cite{salomon,arimondo}, or BECs tunnelling out of a shallow
lattice under gravity \cite{anderson}. Note that at the zone edge
there is a finite probability of Zener tunnelling to the continuum
\cite{note2}. However, we suppress the tunnelling by using a
sufficiently tight lattice, differently from the study performed
in \cite{anderson}. This allows us to keep the atoms oscillating
in the lattice for very long times.

\begin{figure}
\includegraphics[width=8.5cm,clip=]{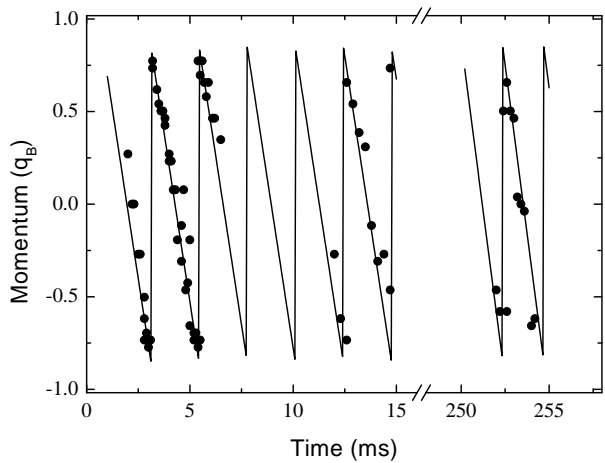} \caption{Bloch
oscillation of the Fermi gas driven by gravity: the peak of the
momentum distribution of the sample scans periodically the first
Brillouin zone of the lattice. More than 100 oscillations can be
followed with large contrast.} \label{fig3}
\end{figure}

\begin{figure}
\includegraphics[width=7cm,clip=]{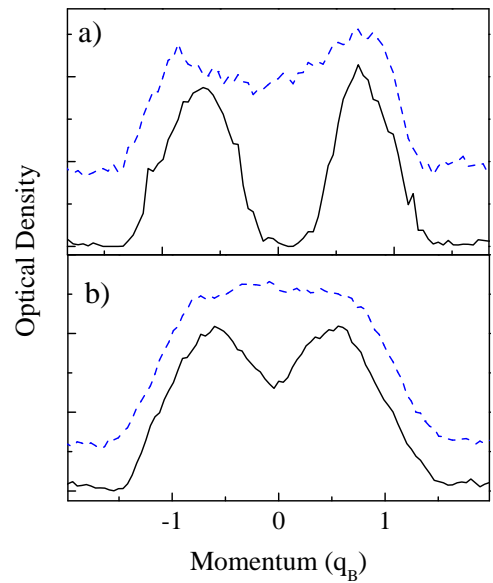} \caption{a) Momentum
distribution of fermions at two different holding times in the
lattice: 1~ms (continuous line) and 252~ms (dashed line). b)
Momentum distribution of bosons at 0.6~ms (continuous line) and
3.8~ms (dashed line). The much faster broadening for bosons is due
to the presence of interactions.} \label{fig4}
\end{figure}

If we follow the vertical position of the peak of the distribution
in Fig.~\ref{fig2}, we get the periodic motion shown in
Fig.~\ref{fig3}, which has the peculiar sawtooth shape expected
for Bloch oscillations \cite{note3}. We can follow the
oscillations for more than 250~ms, that correspond to about 110
Bloch periods, and only at later times the contrast is degraded by
a broadening of the momentum distribution. This is to our
knowledge the longest lived Bloch oscillator observed so far in
all kinds of physical systems. The reduction of contrast is
illustrated in Fig.~\ref{fig4}a. For our parameters ($E_F\approx
E_R$) the initial halfwidth of the wavepacket is $\delta
q\approx0.75q_B$, which fulfills the requirement of a momentum
distribution narrower than the first Brillouin zone of the lattice
to observe the interference. During Bloch oscillations the
distribution broadens steadily and eventually fills completely the
first Brillouin zone.

It is interesting to compare the behavior of fermions and bosons
to study the role of interactions. In our apparatus we can simply
repeat the experiment with a BEC of rubidium atoms. We use a
sample of typically 5$\times10^4$ atoms, at temperatures
$T<$0.6~$T_c$, which is transferred into the lattice with the same
procedure described above for the Fermi gas. The lattice depth is
in the range 2-4~$E_R$ (for rubidium $E_R$=$k_B\times$150~nK). The
phenomenology that we observe is analogous to that found for
fermions: the momentum distribution performs Bloch oscillations
with a period which is now $T_B\approx$1.2~ms because of the
different mass. Two striking differences however appear, as shown
in Fig.~\ref{fig4}b. First of all, at very short times the
momentum width of the BEC is comparable to $q_B$ and therefore
even larger than that of the Fermi gas. This result, that may seem
in contrast with the expectation of a much narrower momentum
distribution for the BEC, is actually the consequence of the
unavoidable conversion of interaction energy into kinetic energy
at the release from the lattice. Also the evolution in the lattice
is affected by interactions, which tend to destroy the
interference between the Wannier-Stark states occupied by the
condensate. A similar phenomenon has already been observed in
presence of gravity \cite{varenna} and in combination with
magnetic traps \cite{njp}. We detect this decoherence as a very
rapid broadening of the momentum distribution, which tends to wash
out the contrast of the Bloch oscillations. As shown in
Fig.~\ref{fig4}, in a lattice with depth $U$=2$E_R$, after
typically 4~ms the momentum distribution fills completely the
Brillouin zone. We have checked that the decay time for the
contrast gets shorter with an increasing lattice depth and radial
confinement, as expected because of the larger density of the
sample. We have measured the longest decay time of about 10~ms,
with a lattice depth of 1.5$E_R$ and an almost absent radial
confinement. In these conditions the lifetime of the sample due to
Zener tunnelling was however comparable to the decay time of the
contrast. We have also repeated the experiment with a cold but
uncondensed cloud of bosons at $T\approx 250$~nK, which again
showed a steady broadening of the distribution due to
interparticle collision. In this case the contrast degraded on a
longer timescale of about 10~ms, which is still much shorter than
the one observed for fermion. This comparison proves the
superiorness of noninteracting fermions with respect to bosons to
perform interferometry with trapped samples.

We now discuss the possible application of this interferometric
scheme to the measurement of forces. From the period of the Bloch
oscillations we can indeed measure the force acting on the atoms
along the lattice as $F$=$2h/T_B\lambda$. As an example, by means
of a nonlinear squares fitting to a sawtooth function, we extract
from the experimental data in Fig.~\ref{fig3} a period
$T_B$=2.32789(22)~ms. Assuming that the only uniform force acting
on the atomic sample is gravity, we determine a local
gravitational acceleration as $g$=9.7372(9)~m/s$^2$. At this level
of sensitivity the relative uncertainty on $g$ is just the same as
on $T_B$, since both $h$ and $m$ are known with a high accuracy
and also $\lambda$ can be accurately determined \cite{lambda}. An
interferometer based on trapped atoms opens the possibility of
probing forces with a high spatial resolution. We note that the
vertical size of the sample in the present experiment is
substantially determined by the initial size in the magnetic trap,
which in principle can be reduced by increasing the vertical
confinement. The minimum possible size is instead set by the
extension of a single Wannier-Stark state, which also corresponds
to the amplitude of the Bloch oscillations in real space. At
$U$=2$E_R$ this amounts to about 4$\mu$m, and decreases further
for increasing depths as 2$\delta/F$, where 2$\delta$ is the width
in energy of the first Bloch band of the lattice.

Clearly, the use of a tight optical lattice to trap the sample
might affect the accuracy of a measurement of forces. In
particular, any axial gradient in the intensity of the lattice
beams will result in an additional force on the sample. In the
experiment we have checked the absence of a dipole force at the
level of our present sensitivity, by repeating the experiment with
a 50\% larger intensity of the lattice beams. This did not produce
a noticeable change of the Bloch period. Since the fermions have a
magnetic moment, the interferometer is sensitive also to magnetic
forces. We actually keep a small homogeneous magnetic field (about
1~G) to avoid spin-flips, which would produce distinguishable
particles. Inhomogeneities in the magnetic field could produce
residual forces. However one can control this effect by repeating
the measurement with two atomic states with different magnetic
moment. Another possibility is to use two different atomic
species. For instance we have compared the values of $g$ measured
in the same conditions with potassium and rubidium, checking the
absence of magnetic forces at the level of 10$^{-3}$ \cite{comp}.

The sensitivity of the present apparatus is limited to 10$^{-4}$
mainly by the 250-ms time interval available for the measurement.
This is much shorter than the characteristic time for $p$-wave
collisions \cite{jin} that we estimate for our sample, which
exceeds 100~s. The main sources of the broadening shown in
Fig.~\ref{fig4}a are presumably intensity and phase noise in the
lattice beams. Also ergodic mixing of the radial and axial
motions, a finite axial curvature of the lattice intensity and a
residual scattering of the lattice photons could contribute to the
observed broadening. A reductions of all these effects, by using
active stabilization of the lattice, a proper beam geometry and a
larger detuning, should allow to extend the observation time to
several seconds, with a corresponding increase of the sensitivity.
The sensitivity can be increased also by using a larger atom
number and/or a longer wavelength for the lattice. Both operations
tend to broaden the momentum distribution with respect to $q_B$:
on the one hand in a Fermi gas the momentum spread increases with
the atom number $N$ according to $\delta q\propto N ^{1/6}$, and
on the other the Brillouin zone shrinks for increasing wavelengths
as $q_B\propto 1/\lambda$. One could however compensate for both
these effects by using a looser radial confinement of the atoms,
which would reduce the momentum spread without affecting the axial
size of the cloud.

In conclusion, we have studied the interference arising from
identical fermions trapped in a vertical optical lattice. The
absence of interactions allows to follow the time-evolution of the
interference for more than 100 periods, whereas in a sample of
bosons this is very rapidly washed out by the interactions.
Interferometry with trapped fermions is promising for a sensitive
determination of forces with high spatial resolution. Possible
applications are the study of forces close to surfaces and at the
sub-millimeter scale, recently motivated by the possibility of new
physics related to gravity \cite{geraci}.

This work was supported by MIUR, by EU under contract
HPRICT1999-00111, and by INFM, PRA ``Photonmatter''. H.O. was
supported by EU under contract HPMF-CT-2002-01958.


\begin{thebibliography}{99}
\bibitem{width} This width is larger than that in the pure magnetic trap, presumably due to a
broadening of the sample during loading in the lattice.
\bibitem{note2} The tunnelling probability is $P$=$\exp(-\lambda \epsilon^2/8\hbar^2 g)$,
where $\epsilon$ in the energy gap between the first and the
second band of the lattice. For $U$=2$E_R$, $P\approx 10^{-7}$ for
potassium atoms.
\bibitem{note3} The distribution peak actually never reaches the edges of the Brillouin
zone, mainly because of a smoothing due to the adiabatic release
from the lattice.
\bibitem{comp} The accuracy of this measurement was limited by the short measurement
time achievable for bosons.
\end{thebibliography}
\end{document}